\documentclass[aps,amsmath,amssymb,superscriptaddress,showpacs,twocolumn]{revtex4}
\usepackage{graphicx}
\usepackage{dcolumn}
\usepackage{bm}

\usepackage{colordvi}
\usepackage{color}
\usepackage{ulem}

\newcommand{\beq}{\begin{equation}}
\newcommand{\eeq}{\end{equation}}
\newcommand{\beqa}{\begin{eqnarray}}
\newcommand{\eeqa}{\end{eqnarray}}

\begin{document}

\title{Spin dynamics in a strongly driven system: very slow Rabi oscillations}

\author{D.V. Khomitsky}
\affiliation{Department of Physics, University of Nizhny Novgorod,
             23 Gagarin Avenue, 603950 Nizhny Novgorod, Russian Federation}

\author{L.V. Gulyaev}
\affiliation{Department of Physics, University of Nizhny Novgorod,
             23 Gagarin Avenue, 603950 Nizhny Novgorod, Russian Federation}

\author{E. Ya. Sherman}
\affiliation{Department of Physical Chemistry, The University of the Basque
Country UPV/EHU, 48080 Bilbao, Spain}
\affiliation{IKERBASQUE Basque Foundation for
Science, Bilbao, 48011 Bizkaia, Spain}

\begin{abstract}
We consider joint effects of tunneling and spin-orbit coupling on driven
by electric field spin dynamics in a double quantum dot with a multi-level resonance scenario.
We demonstrate that tunneling plays the crucial role in the formation of the Rabi-like spin-flip
transitions. In contrast to the linear behavior for weak electric fields,
the spin flip rate becomes much smaller than expected for the two-level model
and shows oscillating dependence on the driving field amplitude in stronger fields.
In addition, the full spin flip is very difficult to achieve in a multi-level resonant system.
These two effects have a similarity with the Zeno effect of slowing down
the dynamics of an observable by its measurement. As a result, spin manipulation by electric field
becomes much less efficient than expected.

\pacs{72.25.Dc,72.25.Pn,73.63.Kv}

\end{abstract}

\maketitle

\section{Introduction}

Fast reliable spin manipulation in quantum dots is one of
the challenges in spintronics and semiconductor-based quantum
information. The design of corresponding gates
can be based on electric dipole spin resonance where the spin-orbit coupling (SOC)
\cite{Rashba1,Nowack,Pioro,Golovach06}
allows on-chip spin manipulation by external electric field
as well as electric read-out of spin states.\cite{Levitov03} Without external driving, SOC effects on localized
in quantum dots electrons are very weak and lead to long spin relaxation times.\cite{Khaetskii01,Fabian05}
The important questions here are how fast can the gate operate,
what limits the manipulation rate, and how efficient
is the spin manipulation in terms of the achievable spin configurations.\cite{GomezLeon11}
It seems that a stronger driving field allows for a faster spin manipulation,
as predicted in a simple Rabi picture of the driven oscillations.
This picture is applicable for a single quantum dot with a parabolic confinement,
where the electron displacement from the equilibrium is linear in the applied electric
field.\cite{Jiang06} However, double quantum dots where tunneling plays the crucial role
for the orbital dynamics, and the corresponding energy scales are different from a single quantum dot,
are more promising for observation of new physics and applications in quantum information
technologies.\cite{Petta05} The tunneling makes the description of the
SOC puzzling since the electron momentum
is not a well-defined quantity at under-the-barrier motion, and the tunneling rate
can become strongly spin-dependent.\cite{Amasha,Stano}
In addition, the double dots provide a possibility to study free and driven interacting qubits.\cite{Shitade11,Nowak}
Here we concentrate on one-dimensional systems attractive for spintronics \cite{quay2010,Pershin04,Malard11}
and building quantum dots \cite{Nadj2010,Nadj-Perge2012,Bringer2011}
and consider spin manipulation in single-electron double quantum dot
\cite{Ulloa06,Sanchez,Wang,Zhu,Wang11,Borhani11} by periodic electric field.

We show that even for a basic quantum system such a
single electron spin, the efficiency and time scale of the
manipulation strongly depend on the electron orbital motion
and, as result, to an unexpected dependence on the external electric field.\cite{diamond_science}
The nonlinearity of the spin and charge dynamics is expected to lead to
unusual consequences on the driven spin behavior.\cite{Rashba11}
In a multilevel system Rabi spin oscillations are slowed down if the field is
sufficiently strong, which challenges efficient spin manipulation.
We restrict ourselves to the single electron dynamics to demonstrate
in the most direct way the nontrivial mutual effect of coordinate and spin motion on the
Rabi oscillations. The slowing of the oscillations down at
high electric fields is a truly unexpected general feature
of a multi-level system compared to the conventional two-level model
and thus can occur in a broad variety of structures.

This paper is organized as follows. In Section II we introduce quantum mechanical description of
electron in a double quantum dot with spin-orbit coupling and magnetic field. Section III
presents the model of driven dynamics. In Section IV we apply the
stroboscopic Floquet approach for the long-time evolution and obtain the properties of Rabi oscillations under
various conditions. Conclusions of this work are given in Section V.

\section{Model, Hamiltonian, and Observables}

\begin{figure}[tbp]
\centering
\includegraphics*[scale=0.3]{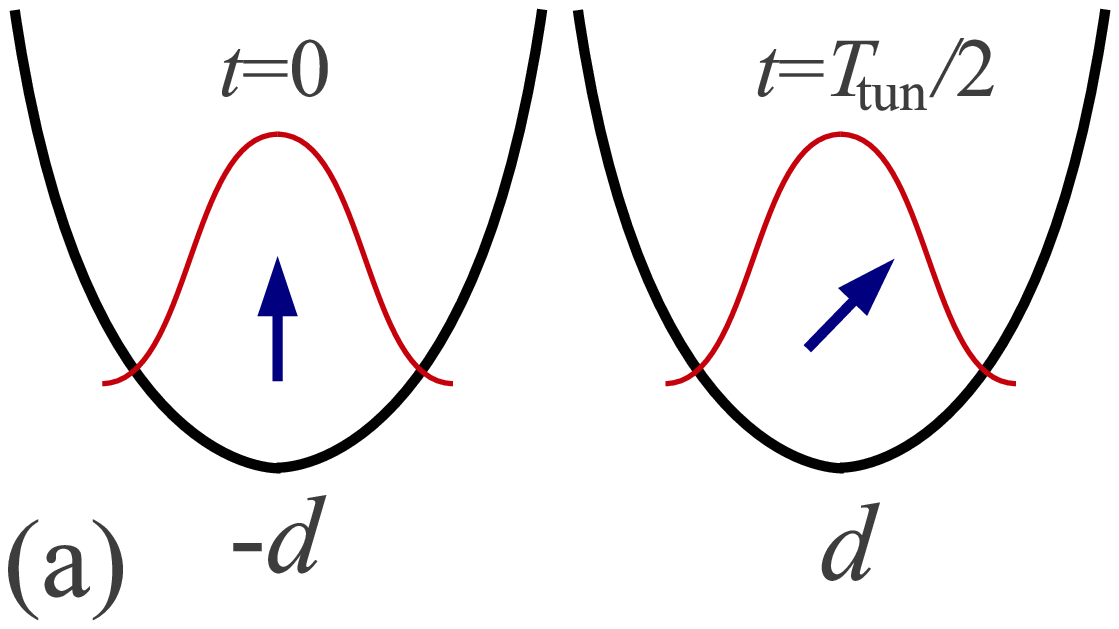}\\
\includegraphics*[scale=0.45]{figure_1.eps}
\caption{(Color online) (a) Qualitative picture of the spin dynamics induced by the interminima tunneling.
(b,c) Free evolution of coordinate (solid line) and spin components $\langle\sigma_x\rangle$ (dashed
line), $\langle\sigma_y\rangle$ (dashed-dot line) for
$B=1.73$ T (b) and $B=6.93$ T (c). The initial state is the even combination of the
states corresponding to the tunneling-split doublet with SOC
taken into account.}
\label{figure1}
\end{figure}

The unperturbed Hamiltonian $H_0={k^{2}}/{2m}+U(x)$ describes
electron in a double quantum dot with the potential (see Fig.\ref{figure1}) \cite{Khomitsky}:
\begin{equation}
U(x)=U_{0}\left[-2\left(\frac{x}{d}\right)^{2}+\left(\frac{x}{d}\right)^{4}\right],
\label{H0}
\end{equation}
where $k=-i\partial/\partial x$ is the momentum operator and $\hbar\equiv 1$.
The minima at $-d$ and $d$ are separated by a barrier of the
height $U_{0}$. In the absence of external fields and SOC
the ground state is split into the doublet of even ($\psi_{\rm g}$) and
odd ($\psi_{\rm u}$) states. The tunneling energy $\Delta E_{g}\ll U_{0}$
determines the tunneling time $T_{\rm tun}=2\pi/\Delta E_{g}$. The Zeeman coupling
to magnetic field $H_{Z}=\Delta_{Z}\sigma_{z}/2$, where $|\Delta_{Z}|$ is the Zeeman splitting.

The SOC has the form:
\begin{equation}
H_{\mathrm{so}}=\left( \alpha _{D}\sigma _{x}+\alpha _{R}\sigma _{y}\right)k,
\end{equation}
where the bulk-originated Dresselhaus $\left(\alpha_{D}\right) $ and
structure-related Rashba $\left(\alpha_{R}\right)$ parameters determine
the strength of SOC.
In the presence of SOC the spatial parity of the eigenstates is  approximate rather than exact
being the qualitative feature of the
coupling linear in the odd $k$-operator, eventually resulting
in the ability of spin manipulation by electric field.

The quantities we are interested in are the spin components:
\begin{equation}
\langle\sigma_{i}(t)\rangle=\int_{-\infty }^{\infty }\sigma _{i}(x,t)dx,
\end{equation}
where $\sigma_{i}(x,t)={\bm \psi}^{\dagger}(x,t)\sigma_{i} {\bm \psi}(x,t)$ is the spin density,
and expectation value of the coordinate
\begin{equation}
\langle x(t)\rangle=\int_{-\infty }^{\infty }x\rho (x,t)dx,
\end{equation}
where $\rho(x,t)={\bm \psi}^{\dagger}(x,t){\bm \psi}(x,t)$, and ${\bm \psi}(x,t)$ is the two-component
electron wavefunction.

For numerical studies we diagonalize exactly the time-independent Hamiltonian
$H_{0}+H_{\mathrm{so}}+{H}_{Z}$ in the truncated spinor basis
$\psi_{n}(x)|\sigma\rangle$, where $\psi_{n}(x)$ are the
eigenfunctions of $H_{0}$ in Eq.(\ref{H0}), and $|\sigma\rangle$
where $\sigma=\pm1$ corresponds to the spin parallel (antiparallel)
to the $z$-axis, find corresponding eigenvalues,
and obtain the new basis set $\left|\psi_{n}\right\rangle $.
We consider below as an example a GaAs-based structure,
where the effective mass is 0.067 of the free electron mass,
with $d=25\sqrt{2}$ nm and $U_{0}=10$ meV.
In the absence of magnetic field the ground state energy is
$E_{1}=3.938$ meV, and the tunneling splitting $\Delta E_{g}=0.092$ meV,
corresponding to the transition frequency close to 23 GHz.
To illustrate the spin dynamics, we consider a moderate external
magnetic field with $\Delta_{Z}=\Delta E_{g}/2$
corresponding to $B=1.73$ T, and a relatively strong magnetic field with $\Delta_{Z}=2\Delta E_{g}$ ($B=6.93$ T)
with the Land\'{e} factor $g=-0.45$. The parameters of the SO coupling are
assumed to be $\alpha_{R}=1.0\cdot 10^{-9}$ eVcm, and $\alpha _{D}=0.3\cdot 10^{-9}$ eVcm,
however, our results can be applied to various double quantum dots with different SOC parameters
and thus have a quite general character.
In particular, the change in the interdot barrier shape and geometry would modify only
quantitatively the system parameters, including the energy levels, spinor
wavefunctions, and, as a result, the resonant driving frequency. The increase in the interdot
distance would decrease the tunneling splitting, making such a system
more sensitive to external influence from phonons, fluctuations in the driving field, etc.

\section{Driven dynamics}

To demonstrate a nontrivial interplay of the tunneling and spin dynamics, we
begin with the coordinate and spin evolution of the electron initially
localized near the $-d$ minimum.  Spin evolution of the state
$\left(\left|\psi_{\rm g}\right>+\left|\psi_{\rm u}\right>\right)\left|1\right>/\sqrt{2}$
can be described approximately analytically taking into account four spin-split lowest
levels and a simpler SOC Hamiltonian $\alpha_{R}\sigma_{y}k$ as
\begin{equation}
\langle\sigma_{x}(t)\rangle= \alpha_{R}\frac{K\Delta E_{g}}{A_{+}A_{-}}\sin (A_{+}t)\sin(A_{-}t)
\end{equation}
where $K=-i\langle\psi_{\rm u}\left| k\right|\psi_{\rm g}\rangle$,
which in the $\Delta E_{g}\ll U_{0}$ limit can be accurately approximated as
$K=md\Delta E_{g}$, and  $A_{\pm}=\sqrt{E_{\pm}^{2}/4+\alpha_{R}^{2}K^{2}}$,
where $E_{\pm}=\Delta E_{g}\pm\Delta_{Z}$.
{ Numerical results for coordinate and spin are shown in Fig.\ref{figure1}.
With the increase in magnetic field, the effect of SOC decreases, leading
to smaller amplitudes of precession, as can be seen from comparison of Fig.\ref{figure1}(b)
and Fig.\ref{figure1}(c). In addition, both the initial state and spin precession axis
change leading to a different phase shift between the observed spin components.}

Next we consider a periodic perturbation by electric field at $t>0$:
\begin{equation}
\mathcal{E}(t)=\mathcal{E}_{0}\sin (\widetilde{\omega}_{Z}t).
\end{equation}
Here $\widetilde{\omega}_{Z}$ is the exact, taking into account SOC,
frequency of the spin-flip transition. For the chosen set of parameters $\widetilde{\omega}_{Z}$
is very close to $\Delta_{Z}.$ The field strength is characterized by
parameter $f$ defined as $e\mathcal{E}_{0}\equiv f\times U_{0}/2d$,
where $e$ is the fundamental charge. For the chosen system parameters, $f=1$
corresponds to the electric field of approximately $1.5\times 10^{3}$ V/cm, similar
to Ref.[\onlinecite{Nowack}].
Here we consider different regimes of the strength and see how the change in
the shape of the quartic potential produced by the field becomes crucially
important for the spin dynamics in two sets of energy levels produced by
magnetic field. We build in the obtained  $|\psi_{n}\rangle$ basis the matrix of the Hamiltonian
$\widetilde{V}=ex\mathcal{E}_{0}\sin(\widetilde{\omega}_{Z}t)$ and study the full
dynamics with the wavefunctions:
\begin{equation}
{\bm \psi}(x,t)=\sum_{n}\xi_{n}(t)e^{-{i}E_{{n}}t}\left|\psi_{{n}}\right\rangle.
\end{equation}
The time dependence of $\xi_{{n}}(t)$ is given by:
\begin{equation}
\frac{d\xi_{{n}}(t)}{dt}={i}e\mathcal{E}(t)\sum_{l}\xi_{l}(t)x_{{ln}}e^{-{i}\left( E_{{l}}-E_{{n}}\right)t},
\label{maineq}
\end{equation}
where $x_{{ln}}\equiv \left\langle \psi _{{l}}\right| \widehat{x}\left|\psi_{{n}}\right\rangle$.
There are two different types of $x_{ln}$: (1) matrix elements of the order of $d$ due to the
different parity of the wavefunctions in the absence of SOC,
and (2) those determined by the SOC strength.
In the weak SOC limit $\left|\Delta_{Z}-\Delta E_{g}\right|\gg\alpha_{R}|K|$,
the SO-determined matrix element of coordinate in the lowest spin-split doublet can be evaluated as

\begin{equation}
x_{\rm so}=2dK\alpha_{R}\frac{\Delta_{Z}}{\Delta_{Z}^{2}-\left(\Delta E_{g}\right)^{2}}.
\end{equation}

In our calculations we assume that the initial state is the ground state of the
full Hamiltonian, that is $\xi_{1}(0)=1$ and $\xi_{n>1}(0)=0$.
The entire driven motion of the system can be approximately
characterized as a superposition of two types of
transitions: resonant ``spin-flip'' transitions with the matrix
element of coordinate determined by the SOC and off-resonant ``spin-conserving''
transitions with a larger matrix element of coordinate.\cite{leakage}
Both types are crucially important
for the understanding of the spin dynamics. { With the estimate $K\approx m\Delta E_{g}d$,
in both cases considered by us ($\Delta_{Z}=E_{g}/2$ and $\Delta_{Z}=2E_{g}$),
we obtain $d\approx 10x_{\rm so}$. As a result, the off-resonant
transitions are not weak compared to the required once.} Throughout the calculation we neglect orbital
and spin relaxation processes assuming that the driving force is sufficiently strong
to prevent the decoherence on the time scale of the spin spin-flip transition.
It is known that the periodic field forms a well-established driven dynamics even in the presence
of damping as long as the level structure is not deeply disturbed by the broadening. For our parameters
it means that one can expect the observation of the predicted results in the currently available semiconductor
structures at temperatures moderately below 1 K.\cite{Nadj2010}

\begin{figure}[tbp]
\centering
\includegraphics*[scale=0.45]{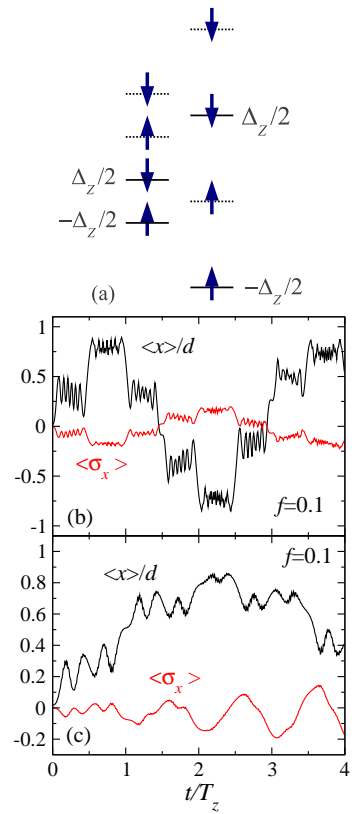}\\
\includegraphics*[scale=0.45]{figure_2.eps}
\caption{(Color online) (a) Scheme of the levels and spin components involved in the Zeeman resonance, $B=1.73$ T (left)
and $B=6.93$ T (right), (b) short-term time-dependent $\langle x\rangle/d$ and $\langle\sigma_{x}\rangle$, as marked
near the plots, induced by external field with $f=0.1$, $B$=1.73 T, { $T_{Z}$=88 ps;} (c) same as in (b) for $B$=6.93 T,
{ $T_{Z}$=22 ps.}}
\label{figure2}
\end{figure}

{ We begin with presentation of the short-time dynamics of coordinate
$\langle x \rangle /d$ and spin $\langle \sigma_{x} \rangle$ for four initial periods of the driving field
(Fig.\ref{figure2}).
These resuts were obtained by the explicit numerical integration of Eq.(\ref{maineq}) with a
time step on the order of $10^{-4}T_{Z}$.}
The other component $\langle\sigma_{z}\rangle$ changes much
slower and will be treated later on a long timescale, which is the primary topic of our interest.
It can be seen in Fig.\ref{figure2} that the fast oscillations are accompanying mainly
the local variations of observables, especially of the spin. { Considerable changes such as
Rabi oscillations of spin can be achieved only after many periods
of the driving field. We will focus on this slow dynamics below.}

\section{Floquet stroboscopic approach}

To { consider the long-term time dependence of the periodically driven system we apply the
Floquet approach \cite{Shirley,Platero,Kohler05,Jiang06,Wu} in the stroboscopic form.}
Here we remind the reader main features of this approach developed in Ref.[\onlinecite{Demikhovskii}].
As the first step, the one-period propagator matrix $\mathbf{U}_{ln}(T_{Z})$ is obtained by a
high-precision numerical integration of the system (\ref{maineq}) at one period of the
driving $T_{Z}=2\pi/\widetilde{\omega}_{Z}$
in the basis of all unperturbed states. For numerically accurate $\mathbf{U}_{ln}(T_{Z})$,
we obtain its eigenvalues $E_{Q}$ which are the quasienergies of the driven system, and the corresponding orthogonal
eigenvectors $A^{Q}_{l}$. As a result,
the one-period propagator $\mathbf{U}_{ln}(T_{Z})$ can be presented as:
\begin{equation}
\mathbf{U}_{ln}(T_{Z})=\sum_{Q} A^{Q}_{l} \left( A^{Q}_{n} \right)^{*} e^{-i E_{Q}T_{Z}}.
\label{uoneper}
\end{equation}
Its $N$-th power obtained by taking into account the orthogonality of the eigenvectors $A^{Q}_{l}$
gives the stroboscopic propagator $\mathbf{U}_{ln}(NT_{Z})$ for $N$ periods as
\begin{equation}
\mathbf{U}_{ln}(NT_{Z})=\sum_{Q} A^{Q}_{l} \left( A^{Q}_{n} \right)^{*} e^{-i E_{Q}NT_{Z}}.
\label{umanyper}
\end{equation}
For any integer $N$ the system state is
given by $\left|\Psi(NT_{Z})\right\rangle=\mathbf{U}_{ln}(NT_{Z})\left|\Psi(0)\right\rangle $.
{ The similarity of Eq.(\ref{uoneper}) for a single-period propagator and Eq.(\ref{umanyper})
for any $N\ge1$ is a highly nontrivial fact demonstrating that
${U}_{ln}(NT_{Z})={U}_{ln}^{N}(T_{Z})$ can be simply expressed by the right-hand-side
in Eq.(\ref{umanyper}).}
The stroboscopic approach allows us to study very
accurately the long-time evolution since the $N$-period propagator (\ref{umanyper}) is constructed
explicitly in a finite algebraic form. Although this propagator describes the dynamics
exactly, it allows to watch only the stroboscopic evolution rather than the entire
continuous one. However, if we are interested in slowly evolving phenomena such as
chaos development \cite{Demikhovskii} and Rabi oscillations which occur here on many periods
of the driving field, the stroboscopic approach is fully justified and highly efficient.
{The experiment \cite{Nowack} uses stroboscopic approach with the intervals
on the order of 100 ns to measure the slow dynamics of the driven electron spin.}

\begin{figure}[tbp]
\centering
\includegraphics*[scale=0.45]{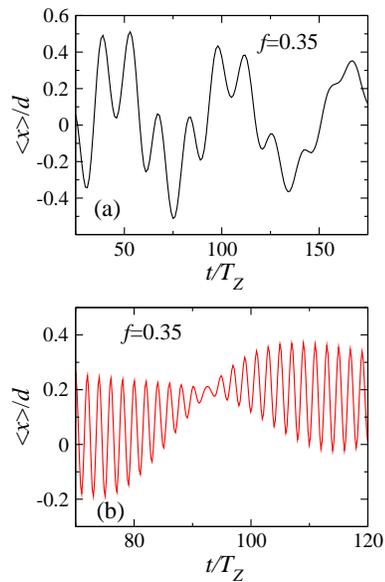}
\caption{(Color online) Stroboscopic time dependence of $\langle x\rangle/d$ for field $f=0.35$ in a given time window:
(a)  $B=1.73$ T, (b) $B=6.93$ T. {Solid lines serve only as a guide for the eye.}}
\label{figure3}
\end{figure}

\begin{figure}[tbp]
\centering
\includegraphics*[scale=0.45]{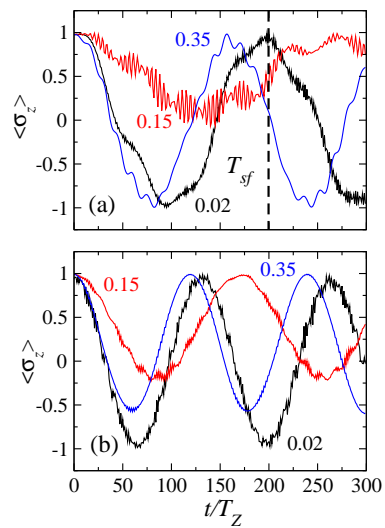}
\caption{(Color online) Stroboscopic time dependence of $\langle\sigma_{z}(NT_{Z})\rangle$ for different external
driving fields $f$ marked near the plots for two different magnetic fields: (a)  $B=1.73$ T, vertical dashed line marks
the operational definition of the spin-flip period, (b) same as in (a) for $B=6.93$ T.}
\label{figure4}
\end{figure}

\begin{figure}[tbp]
\centering
\includegraphics*[scale=0.45]{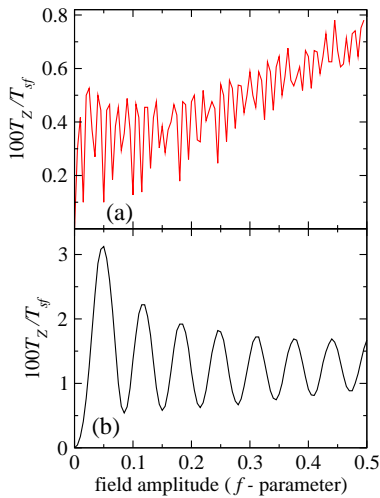}
\caption{(Color online) Nonlinear dependence of the Rabi spin-flip frequency on the electric field amplitude in
multilevel system for two different magnetic fields (a) $B=1.73$ T, the characteristic spin-flip rate of the order
of 50 MHz, (b) $B=6.93$ T, the characteristic spin-flip rate of the order of 400 MHz.}
\label{figure5}
\end{figure}

{ The results of calculations of electron displacement at discrete times $NT_{Z}$ are presented in Fig.\ref{figure3}.
As one can see in Fig.\ref{figure3}, the time dependence of the
displacement becomes strongly nonperiodic with the typical values being considerably less than $d$.
That is in a strong electric field the electron probability density redistribution between the dots
is not complete. The motion can be qualitatively analyzed in the pseudospin model
of the charge dynamics,\cite{Rashba11} where the tunneling splitting is described by the $\sigma_{z}$ matrix,
and the driving field is coupled to the $\sigma_{x}-$matrix. The decrease in the electron displacement
with the increase in the electric field can be viewed as a suppressed spin precession in a strong periodic field
or as a coherent destruction of tunneling.\cite{Platero}
This nonperiodic behavior and decrease in the displacement eventually result in a less efficient spin driving.}
It should be mentioned that the fast oscillations in Fig.\ref{figure2} which are in general absent
in Fig.\ref{figure3} reflect the difference between the continuous time scale in the former Figure and
the stroboscopic Floquet times $NT_{Z}$ in the latter one. { Figure \ref{figure3} clearly illustrates
the role of the spin in the orbital dynamics: the curves in Fig.\ref{figure3}(a) and \ref{figure3}(b)
are very different. Tracking of the system at stroboscopic times $NT_{Z}$ may not allow seeing the complete
fast orbital dynamics, thus masking some details. As a result,
there is no simple way to describe this stroboscopic picture directly in terms
of the Hamiltonian parameters.}

The slow long-term spin dynamics is presented in Fig.\ref{figure4}. Here the "unit of time" $T_{Z}$ is short enough and the time dependence of
$\langle\sigma_{z}\rangle$ is accurately described by the stroboscopic approach.
Since the spin dynamics is not strictly periodical and full spin flips do not always appear
in this system, we use the operational definition of the ``spin-flip'' time $T_{sf}$:
spin flip occurs when spin component shows a broad minimum albeit accompanied by fast oscillations
(see in Fig.\ref{figure4}(a)). The fast dynamics in the spin-flip doublet shown in Fig.\ref{figure4}
becomes slow with the field increase as a result of a weaker
effective coupling of the states with different parity. The resulting spin
behavior, arising solely due to the SOC, is shown in Fig.\ref{figure4}.
The Rabi frequency for the spin-flip is smaller for some higher
values of $f$ (which we vary through Fig.\ref{figure3}-Fig.\ref{figure4}) than for some weaker values of $f$
in contrast to what can be expected for the weak fields
employed, e.g. in the experiments \cite{Nowack}, being a manifestation of the generally
nonmonotonous behavior of the Rabi frequency on the electric field amplitude.
In addition, in contrast to the simple Rabi oscillations, the
flips become incomplete, with $\left<\sigma_{z}(t)\right>=-1$ never reached. These two
qualitative effects are the results of the enhanced electron tunneling
between the potential minima: the spin precession in the driven interminima motion
establishes corresponding spin dynamics and prevents the electric field to flip
the spin efficiently. This effect makes a qualitative difference to the model
of Ref.\cite{Nowack}, where electron is assumed to be always located in the orbital ground state near
the minimum of the potential formed by the parabolic confinement and weak external electric field.

To present a broader outlook onto the dependence of
the spin flip rate on the driving field, we plot in Fig.\ref{figure5} the spin flip rate
for $B=1.73$ T and $B=6.93$ T. In contrast to the linear dependence
for a conventional two-level Rabi resonance formula, one can see a strongly
different much more complicated non-monotonous
dependence in a multi-level structure, especially at high fields. The regime in Fig.\ref{figure5}(a)
shows more irregularities since all four lowest
levels are equidistant (Fig.\ref{figure2}(a))
and involved in the resonance while in Fig.\ref{figure5}(b) more regular dependence is observed,
reflecting a simpler nature of the resonances here.

We would like to mention here that the observed slowing down
of spin dynamics can be seen on a more general ground,
not restricted to the exact form of Eq.(\ref{maineq}), as
the Zeno effect of freezing evolution of a measured quantity.\cite{Streed,Echanobe,Sokolovski10}
Indeed, the operator $-i\sigma_i\partial/\partial x$ makes the orbital dynamics spin-dependent,
and, as a result, performs the measurement of the $\sigma_i$ component \cite{Sokolovski11,Allahverdyan}
in the sense of von Neumann procedure. This can be seen in the evolution
of a two-component wave function:\cite{Sokolovski11}
\begin{eqnarray}
&&e^{-\alpha t\sigma_z{\partial}/{\partial x}}\phi(x)
\left(\zeta_{1}\left|1\right>+\zeta_{-1}\left|-1\right>\right)
= \nonumber \\
&&\phi(x-\alpha t)\zeta_{1}\left|1\right>+\phi(x+\alpha t)\zeta_{-1}\left|-1\right>,
\end{eqnarray}
where we took $i=z$ as an example,
$\zeta_{1}$ and $\zeta_{-1}$ correspond to $\pm1$ eigenvalues of $\sigma_{z}$, respectively,
and $\alpha$ is the coupling constant. The SOC thus The SOC thus entangles the orbital and spin motion, destroys
the coherent superposition of spin-up and spin-down states, and performs the von Neumann-like spin measurement
by mapping spin state on the electron position.
This von Neumann measurement, is, however, different from the experimental measurement procedure applied,
e.g. in Ref.[\onlinecite{Nowack}].  The spin-orbit coupling coupling drives the coherent superposition of different
spin components and at the same time, by constant strong measurement, destroys it leading to a slow spin dynamics.

\section{Conclusions}

We have considered the interplay between the tunneling and
spin-orbit coupling in a driven by an external electric
field one-dimensional single-electron double quantum dot.
In the regime of the electric dipole spin resonance, where the electric
field frequency exactly matches the Zeeman transition, the complex interplay of these
mechanisms results in two unexpected effects. The first effect is the
nonmonotonous change in the Rabi spin oscillations
frequency with the electric field amplitude. The Rabi oscillations  become much slower
than expected for a two-level system. The second effect is the incomplete Rabi spin flips.
{ This behavior results from the fact that the interminima motion establishes a competing
spin dynamics, leading to the physics somewhat similar to the Zeno effect, preventing a fast
change in a measured quantity.}
These results indicating the slowdown and nonlinearity of the spin resonance in multilevel systems can be useful
for pointing out certain fundamental challenges for the future experimental and spintronics device applications
of  phenomena based on spins in double quantum dots.

\section{Acknowledgements}  D.V.K. is supported by the RNP Program of Ministry of Education and Science RF,
and by the RFBR (Grants No. 11-02-00960a, 11-02-97039/Regional).  This work of EYS was supported by the MCINN of Spain
Grant FIS2009-12773-C02-01, by "Grupos Consolidados UPV/EHU del Gobierno Vasco" Grant IT-472-10,
and by the UPV/EHU under program UFI 11/55.

\end{document}